\documentclass{aastex631}

\usepackage{graphicx}

\usepackage{colortbl}
\usepackage{xcolor}
\usepackage{float}
\usepackage{subfigure}
\usepackage{multirow}
\usepackage{booktabs}

\usepackage{times}
\usepackage{tablefootnote}
\usepackage{overpic}
\usepackage{threeparttable}
\usepackage{hyperref}

\usepackage{amsmath} 

\begin{document}

\title{Is the Gum Nebula an Important Interstellar Scattering Disk of Background Pulsars?}

\correspondingauthor{Zhen Yan, Zhiqiang Shen}
\email{yanzhen@shao.ac.cn, zshen@shao.ac.cn}

\author{Rui Wang}
\affiliation{Shanghai Astronomical Observatory,Chinese Academy of Sciences, Shanghai,200030, People's Republic of China;}
\affiliation{School of Astronomy and Space Sciences, University of Chinese Academy of Sciences, Beijing,100049, People's Republic of China;}

\author{Zhen Yan}
\affiliation{Shanghai Astronomical Observatory,Chinese Academy of Sciences, Shanghai,200030, People's Republic of China;}
\affiliation{School of Astronomy and Space Sciences, University of Chinese Academy of Sciences, Beijing,100049, People's Republic of China;}
\affiliation{Key Laboratory of Radio Astronomy and Technology, Chinese Academy of Sciences, Beijing,100101, People's Republic of China;}

\author{Zhiqiang Shen}
\affiliation{Shanghai Astronomical Observatory,Chinese Academy of Sciences, Shanghai,200030, People's Republic of China;}
\affiliation{School of Astronomy and Space Sciences, University of Chinese Academy of Sciences, Beijing,100049, People's Republic of China;}
\affiliation{Key Laboratory of Radio Astronomy and Technology, Chinese Academy of Sciences, Beijing,100101, People's Republic of China;}
\affiliation{School of Physical Science and Technology,ShanghaiTech University, Shanghai,201210, People's Republic of China;}

\author{KeJia Lee}
\affiliation{Department of Astronomy, School of Physics, Peking University, Beijing,100871, People's Republic of China;}
\affiliation{Kavli Institute for Astronomy and Astrophysics, Peking University, Beijing,100871, People's Republic of China;}

\author{Yajun Wu}
\affiliation{Shanghai Astronomical Observatory,Chinese Academy of Sciences, Shanghai,200030, People's Republic of China;}
\affiliation{Key Laboratory of Radio Astronomy and Technology, Chinese Academy of Sciences, Beijing,100101, People's Republic of China;}

\author{Rongbing Zhao}
\affiliation{Shanghai Astronomical Observatory,Chinese Academy of Sciences, Shanghai,200030, People's Republic of China;}
\affiliation{Key Laboratory of Radio Astronomy and Technology, Chinese Academy of Sciences, Beijing,100101, People's Republic of China;}

\author{Zhipeng Huang}
\affiliation{Shanghai Astronomical Observatory,Chinese Academy of Sciences, Shanghai,200030, People's Republic of China;}
\affiliation{School of Astronomy and Space Sciences, University of Chinese Academy of Sciences, Beijing,100049, People's Republic of China;}
\affiliation{School of Physical Science and Technology,ShanghaiTech University, Shanghai,201210, People's Republic of China;}
\affiliation{Department of Physics and Astronomy, Hubei University of Education,Wuhan,430205,People's Republic of China}

\author{Xiaowei Wang}
\affiliation{Shanghai Astronomical Observatory,Chinese Academy of Sciences, Shanghai,200030, People's Republic of China;}
\affiliation{School of Astronomy and Space Sciences, University of Chinese Academy of Sciences, Beijing,100049, People's Republic of China;}

\author{Jie Liu}
\affiliation{Shanghai Astronomical Observatory,Chinese Academy of Sciences, Shanghai,200030, People's Republic of China;}



\begin{abstract}
The Gum Nebula is a faint supernova remnant extending about 40 degrees across the southern sky, potentially affecting tens of background pulsars. Though the view that the Gum Nebula acts as a potential scattering screen for background pulsars has been recurrently mentioned over the past five decades, it has not been directly confirmed. We chose the strong background pulsar PSR~B0740$-$28 as a probe and monitored its diffractive interstellar scintillation (DISS) at 2.25~$\&$~8.60~GHz simultaneously for about two years using the Shanghai Tian Ma Radio Telescope (TMRT). DISS was detected at both frequencies and quantified by two-dimensional autocorrelation analysis. We calculated their scattering spectral index $\alpha$ and found that 9/21 of the observations followed the theoretical predictions, while 4/21 of them clearly showed $\alpha < 4$. This finding provides strong support for anomalous scattering along the pulsar line of sight, due to the large frequency lever arm and the simultaneous features of our dual-frequency observations. In comparison to the 2.25~GHz observations, scintillation arcs were observed in 10/21 of the secondary spectrum plots for 8.60~GHz observations. Consequently, the highest frequency record for pulsar scintillation arc detection was updated to 8.60~GHz. Our fitting results were the most direct evidence for the view that the Gum Nebula acts as the scattering screen for background pulsars, because both the distance ($245^{+69}_{-72}$~pc) and transverse speed ($22.4^{+4.1}_{-4.2}$~km/s) of the scintillation screen are comparable with related parameters of the Gum Nebula. Our findings indicated that anisotropic scattering provides a superior explanation for the annual modulation of scintillation arcs than isotropic scattering. Additionally, the orientation of its long axis was also fitted.
\end{abstract}



\section{Introduction}

\begin{figure*}[ht]       
\subfigure{\begin{minipage}[t]{0.98\linewidth}
\centering       
\includegraphics[height=10.6cm,width=16cm]{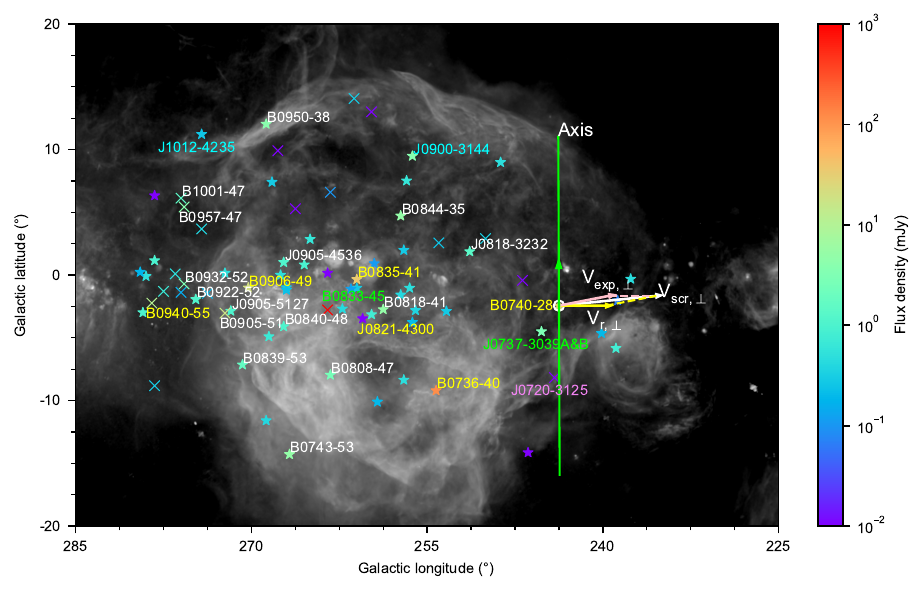}       
\end{minipage}}
\vspace{-0.5cm}
\caption{Plots about pulsars inside ($\times$ marks) and behind ($\star$ marks) the Gum Nebula. The parameters of pulsars and the H$\alpha$ map \citep{Fin03} of the Gum Nebula are obtained from the ATNF Pulsar Catalogue (PSRCAT) \citep{mht05} and the Southern H-Alpha Sky Survey Atlas \citep{gmr01}, respectively. The names of pulsars (with $S_{\rm 1400}>1$~mJy) are labeled on the figure with different colors based on methods of obtaining their distances. The names of the pulsars whose distances are estimated by the YMW16 Galactic electron density model \citep{ymw17} are shown in white, while those with independent distance measurements based on HI absorption lines \citep{vjw12, rdg95}, Very Long Baseline Interferometry \citep{dbt09, dlr03}, pulsar timing \citep{dcl16, gf24}, and optical parallax of companion star \citep{kva07} are labeled with yellow, lime, cyan, and violet, respectively. See section~\ref{sec:discuss} for more information about Axis (lime line), $\vec{V}_{\rm scr, \bot}$ (white vector), $\vec{V}_{\rm exp, \bot}$ (pink vector), and $\vec{V}_{\rm r, \bot}$ (yellow vector). }
\label{fig:gum}
\end{figure*}

The Gum Nebula is the closest supernova remnant, located about 350~pc away from the Earth and spanning across about 40$^{\rm \circ}$ in the southern sky \citep{sko22}. It is so large and faint that it is easily lost in the din of a bright and complex background. In addition to the $H\alpha$ image of the Gum Nebula shown in Figure~\ref{fig:gum}, pulsars that may be located inside and behind this nebula are also plotted with crossings ($\times$) and stars ($\star$) markers, colored according to their flux densities at 1400~MHz ($S_{\rm 1400}$), respectively. The location of the pulsars inside or behind the Gum Nebula is determined by comparing their distances with the spherical front and back shells of the nebula. As early as 1974, Backer suggested that the strong scattering of the Vela pulsar (PSR~B0833$-$45) could be attributed to the Gum Nebula \citep{bac74}. Based on the observation that the interstellar scintillation (ISS) speeds of PSR~B0835$-$41 and PSR~B0740$-$28 were much lower than their speeds measured with other methods, researchers also pointed out that the Gum Nebula might serve as the scintillation screen \citep{jnk98}. Although enhanced scattering phenomena of more pulsars behind the Gum Nebula were observed, it was still insufficient to definitively conclude that they were caused by this nebula, as the interstellar scattering observations cannot directly determine the location of the scattering screen \citep{mr01}. Furthermore, it was mentioned once more that the strange scintillation of the Vela pulsar should be caused by a hot bubble limited by the Gum Nebula \citep{zew05}. Overall, the perspective that the Gum Nebula acts as a potential scattering screen for background pulsars has been recurrently discussed over the past five decades, but it has not been confirmed yet.

It is theoretically possible to determine the position of the scattering screen using pulsar ISS observations \citep{sti06, wms04}. For example, the H~II regions and supernova remnants have been studied using ISS observations of background pulsars \citep{mma22, yzm21}. Shortly after the discovery of the first pulsar, diffractive interstellar scintillation (DISS) was proposed to explain flux density fluctuations of pulsars on short timescales ($\sim$~min) and narrow bandwidths ($\sim$~kHz $-$ MHz), which were named the scintillation timescale ($\Delta\tau_{\rm d}$) and the decorrelation bandwidth ($\Delta\nu_{\rm d}$), respectively \citep{sch68, ric69}. As research progressed, longer-term variations in pulsar flux density (tens of days or longer) were also discovered and explained by refractive interstellar scintillation (RISS) \citep{rcb84}. DISS and RISS are essentially caused by the diffraction and refraction of irregularities within the interstellar medium (ISM), with typical scales of $10^{\rm 6}-10^{\rm 8}$~m and $10^{\rm 10}-10^{\rm 12}$~m, respectively.

In addition, the ISS observation can also be used to study the ISM fluctuation in the corresponding direction \citep{sch68, ric69, Nar92}. For scales ($L$) ranging from the inner-scale ($L_{\rm in}$) to the outer-scale ($L_{\rm out}$) cutoffs, the electron density fluctuation of the ISM can normally be described with a Kolmogorov-like spectrum expressed as:
\begin{align}
P_{\rm 3n}(q)&=C_{\rm N}^{\rm 2}(q^{\rm 2}+L_{\rm out}^{\rm -2})^{\rm -\beta/2}*exp[-q^{\rm 2}*4L_{\rm in}^{\rm 2}],
\label{eq:ismspec}
\end{align}
where $q=2\pi/L$. $\beta$ and $C_{\rm N}^{\rm 2}$ are the spectral index and the fluctuation coefficient, respectively \citep{ars95}. When $L_{\rm in} << 1/q << L_{\rm out}$, Eq.~\ref{eq:ismspec} can be simply expressed as $P_{\rm 3n}(q)\approx C_{\rm N}^{\rm 2}*q^{\rm -\beta}$. The possible range of $\beta$ is from 3 to 5, and $\beta=11/3$ is associated with the well-known Kolmogorov spectrum. While some views reject $\beta>4$ \citep{rl90, ars95}, a series of anomalies suggested the presence of steeper power-law spectra ($\beta\ge4$), such as fringing in dynamic spectra, which implies the occurrence of multiple imaging \citep{cw86,rlg97} and extreme scattering events \citep{lrc98,lkm01}. As ISS parameters (like $\Delta\tau_{\rm d}$ and $\Delta\nu_{\rm d}$) change with observation frequency ($\nu$) following power-law indices related to $\beta$, it is possible to directly fit $\beta$ using multi-frequency ISS observations. In order to obtain a reliable $\beta$ result, multi-frequency ISS observations are required to be arranged simultaneously. By doing so, it can be ensured that the ISS is caused by the same clumps of ISM. However, there have been few simultaneous multi-frequency ISS observations of pulsars, primarily due to the following factors: Firstly, the flux density of pulsars tends to decrease rapidly with increasing $\nu$, following a steep power law (with an average spectral index around $-$1.5). Secondly, the scintillation strength ($u$) also weakens as $\nu$ increases, following the relation of $u=\sqrt{2\nu/\Delta\nu_{\rm d}}\propto \nu^{\rm -1/2-2/(\beta-2)}$ \citep{Ric90}. Thirdly, few telescopes supported simultaneous multi-frequency pulsar observations in the past. Building on the technical advancements in high-sensitivity wide-band observations, some researchers have tried to obtain analogous simultaneous multi-frequency ISS observations by dividing the total bandwidth into several sub-bands \citep{tdc24}.

Based on the flux densities and positions of background pulsars of the Gum Nebula (see Figure~\ref{fig:gum}), we selected PSR~B0740$-$28 as the probe to clarify the potential effects of the Gum Nebula and study electron fluctuation along this direction. PSR~B0740$-$28 is a relatively strong pulsar with $S_{\rm 1400}=26$~mJy and a spectral index of about $-2.0$ \citep{lyl95, jvk18}. Although there are some stronger pulsars in this direction, the lower limits of their estimated distances are comparable with that of the Gum Nebula. In comparison, PSR~B0740$-$28 is located at a distance of $D=2.0^{+1.0}_{-0.8}$~kpc \citep{vjw12} and moves with a proper motion of [$-$29.0, $-$3.7]~mas/yr in the Ecliptic coordinate system \citep{fgm97}. Previous DISS observation results of PSR~B0740$-$28 ranging from 628~MHz to 10.7~GHz are presented in Table~\ref{tab:preresults} \citep{mss96, jnk98, whh18}. Although the values of $\Delta\tau_{\rm d}$ were successfully derived in all of these observations, the values of $\Delta\nu_{\rm d}$ were occasionally obtained due to insufficient frequency resolution ($\Delta\nu_{\rm 0}$) or even only recording the total power. In comparison, we conducted long-term observations of this pulsar at 2.25~$\&$~8.60~GHz in a simultaneous manner with the Shanghai Tian Ma Radio Telescope (TMRT) \citep{ysw2015}. In the following part of this paper, we will provide detailed information about our simultaneous 2.25~$\&$~8.60~GHz observations on PSR~B0740$-$28 in Section~2. Relevant data analysis will be presented in Section~3, and the observation results are listed in Section~4. Finally, discussions and conclusions about our results will be given.

\begin{figure*}[htb]
\centering  
	\subfigure{\begin{minipage}[t]{0.98\linewidth}      
\includegraphics[height=11.0cm,width=17cm]{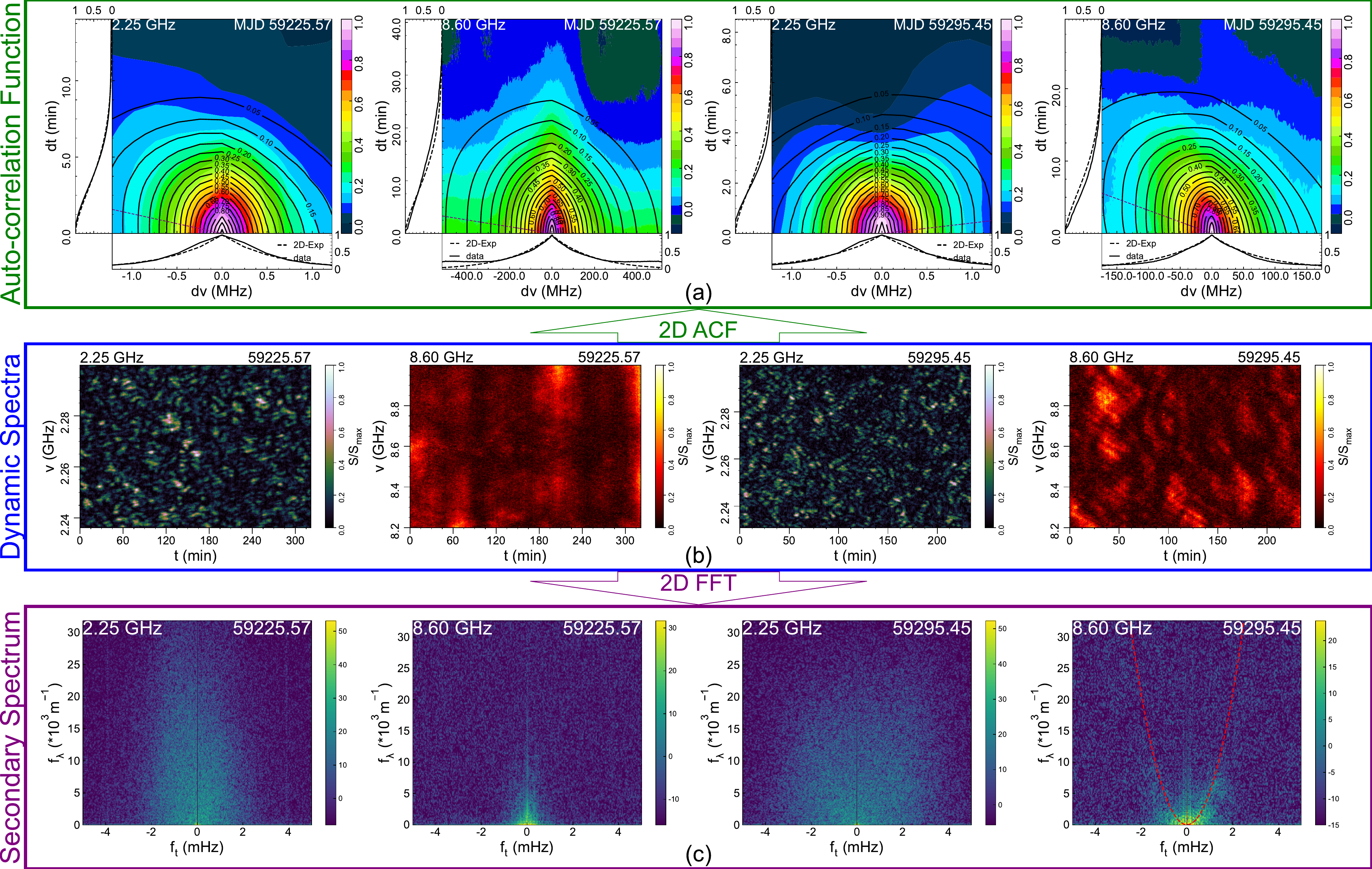}  
\end{minipage}}
\vspace{0.5cm}
\caption{Plots of two pairs of samples from 2.25~$\&$~8.60~GHz observations on MJDs~59225.57 and 59295.45. (a) 2D-ACF plots: In each central sub-panel, the normalized 2D-ACF results are shown with color-filled contours (scaled with the color bar on the right), and the 2D-Exponential form function fitting results are shown with black-line contours, with the corresponding major axis labeled with the dashed line. In each bottom and left sub-panel, the 1D-ACF results at zero time and frequency lag are shown with solid curves, while the curves for the best-fitting $\Delta\nu_{\rm d}$ and $\Delta\tau_{\rm d}$ are plotted with dashed lines correspondingly. (b) Dynamic spectrum plots: The integration time of dynamic spectra is 30~s, and their frequency resolutions are 1.95~MHz at 8.60~GHz and 0.24~MHz at 2.25~GHz. The brightness of each pixel is linearly scaled with a color bar on the right. (c) Secondary spectrum plots: The red dashed curve in the rightmost sub-panel is the  best-fitting arc structure.}
\label{fig:opra}
\end{figure*}

\begin{table*}
\caption{Historical ISS observation information and results of PSR~B0740$-$28: Modified Julian Date (MJD), frequency range (FR), frequency resolution ($\Delta\nu_{\rm 0}$), observation length ($T$), decorrelation bandwidth ($\Delta\nu_{\rm d}$), decorrelation time ($\Delta\tau_{\rm d}$), drift rate ($dt/d\nu$), reduced ISS speed ($V_{\rm iss, 0}$), telescope, and reference.}
\renewcommand\arraystretch{1}
{
\begin{tabular*}{1\linewidth}{ccccccccccc}
\toprule
\multirow{2}{*}{MJD}   & FR       &  $\Delta\nu_{\rm 0}$    & T     & $\Delta\nu_{\rm d}$     & $\Delta\tau_{\rm d}$    & $dt/d\nu$  & $V_{\rm iss,0}$  & Telescope &Reference   \\
 & (GHz)   & (MHz) & (s)  & (MHz) & (min) & (s/MHz) & (km/s)  &&& \\ 
\midrule
$-$& 0.628$\sim$0.692  & 0.125 &{10872}  & $-$ & 0.97& $-$&$-$   & Parkes & \citep{jnk98} \\	
57368.69  & 2.2$-$2.3 & 0.58 &{10080} & $1.1\pm0.1$ &$ 5.2\pm0.2$&$-3.0\pm7.2$ &$81\pm5$    &  Jiamusi  &   \citep{whh18}  \\	
57413.54  & 2.2$-$2.3  & 1.17 &{11160}  &$ 1.1\pm0.1 $ &$ 2.9\pm0.2$  &$12.0\pm8.4$      &$146\pm8$    &      Jiamusi   &   \citep{whh18}      \\	
58058.79  & 2.2$-$2.3  & 0.58 &{9360}  & $0.6\pm0.1$ & $3.0\pm0.2  $&$-32.4\pm11.4 $   &$104\pm9$    & Jiamusi    &  \citep{whh18}    \\	
49081 & 4.5$-$5.0  & $-$&{5700}  & $-$& $3.5\pm1$  &    $-$&  $-$ & Effelsberg &  \citep{mss96} \\
-& 4.64$-$4.96  & 5 &{8028}  & 8.83 & 10.6  &   $-$  & 56.0  & Parkes &\citep{jnk98}   \\
-& 8.24$-$8.56  & 5 &{3852}  & 40.0 & 37.7  &$-$& 18.1  & Parkes &\citep{jnk98} \\	
48499 & 10.4$-$10.7  & $-$&{3600} & $-$& $5\pm1$ &$-$&$-$  & Effelsberg  &\citep{mss96} \\
\bottomrule
\end{tabular*}}
\label{tab:preresults}
\end{table*}

\section{Observations}
The TMRT is a fully-steerable antenna with a diameter of 65~m located in the suburban Songjiang district of Shanghai City. Our simultaneous dual-frequency ISS observations of PSR~B0740$-$28 employed the parallel capabilities of the 2.25~$\&$~8.60 GHz dual-frequency receiver and digital back-end system (DIBAS), which is composed of three pairs of parallel units for data sampling, processing, and recording. The effective bandwidths for the receiver at 2.25~$\&$~8.60~GHz are 100 and 800~MHz, respectively \citep{ysw2015,ysw18}. In our observations, we chose to use the online incoherent de-dispersion and folding mode (sub-integration length 30~s) of the DIBAS. During the de-dispersion process, the total bandwidth at 2.25~$\&$~8.60~GHz was divided into sub-channels with a typical width of 0.24 (or 0.97) and 0.49 (or 1.95)~MHz, respectively. For the convenience of subsequent data processing with widely-used software, the observation data were finally written in the standard PSRFITS format \citep{hvm04}. For the 8.60~GHz observation data with the frequency resolution $\Delta\nu_{\rm 0} = 0.49$~MHz, we merged four adjacent channels to achieve higher sensitivity using the \texttt{PSRCHIVE} \citep{hvm04,swd12} \footnote{\label{foot:psr}\url{http://psrchive.sourceforge.net/}}, and $\Delta\nu_{\rm 0}$ was reduced to 1.95~MHz, while $\Delta\nu_{\rm 0}$ at 2.25~GHz remained the same as DIBAS wrote out. We sequentially list the Modified Julian Day (MJD), frequency $\nu$, final-used $\Delta\nu_{\rm 0}$, and length ($T$) of our observations in columns 1$-$4 of Table.~\ref{tab:ourresults}.

\section{Data analysis and results}
Firstly, the band-pass calibration was done using the standard calibrators (3C~286 or 3C~295). Secondly, radio frequency interference (RFI) was removed with an automatic script (\texttt{rfihunter}\footnote{\label{foot:rfi}\url{https://github.com/mshamohammadi/rfihunter}}) and an interactive command of the \texttt{PSRCHIVE} (\textsl{pazi}). Thirdly, we generated the temporary dynamic spectrum using the \texttt{psrflux} of \texttt{PSRCHIVE}. Because some isolated abnormal (or deleted) pixels remained in the time and/or frequency domains, linear interpolation was applied to make them more continuous.

\renewcommand\arraystretch{1}
\begin{table*}[!t]
\caption{Information about our 2.25~$\&$~8.60~GHz ISS observations on PSR~B0740$-$28 and related results: MJD, observation frequency ($\nu$), $\Delta\nu_{\rm 0}$, $T$, $\Delta\nu_{\rm d}$, $\Delta\tau_{\rm d}$, $dt/d\nu$, curvature of arc ($\eta$), the scattering index ($\alpha$), the fluctuation index ($\beta$), $V_{\rm iss, 0}$, refractive steering angle ($\theta_{\rm r}$), the pulse scatter broadening timescale at 1~GHz ($t_{\rm d, 1~GHz}$), and the modulation index $m$.}
\resizebox{\linewidth}{!}{
\begin{tabular}{*{13}{>{}c}*{1}{>{}c}}
\toprule
\multirow{2}{*}{MJD} &$\nu$&  $\Delta\nu_{\rm 0}$    & $T$ & $\Delta\nu_{\rm d}$  & $\Delta\tau_{\rm d}$ & $dt/d\nu$   &$\eta$     & \multirow{2}{*}{$\alpha$}   &\multirow{2}{*}{$\beta$}  & $V_{\rm iss, 0}$ &$\theta_r$     & $t_{\rm d, 1~GHz}$&\multirow{2}{*}{$m$}        \\
&(GHz) & (MHz) & (s) & (MHz) & (min) & (s/MHz) &${\mathbf{\rm (m^{-1}mHz^{-2})}}$&    & & (km/s)  &($\mu$as)       &$(\mu$s)   &         \\
\midrule
&2.25 & 0.97 &  & $ 0.63\pm0.02\downarrow$ &$ 3.59\pm0.10 $&$ -28.0\pm2.7 $&  & & &$89.3\pm3.0\downarrow$& $-7.1\pm0.7\downarrow$& &0.78\\
\multirow{-2}{*}{59168.83} &8.60& 1.95 &\multirow{-2}{*}{10673} &$88\pm25 $&$ 27\pm5 $&$ 0.25\pm0.12$&$- $&\multirow{-2}{*}{$-$}  &\multirow{-2}{*}{$-$} &$36\pm9 $&$ 0.10\pm0.05 $&\multirow{-2}{*}{$-$}&1.00\\
\midrule
&2.25 & 0.97 &  & ${0.66\pm0.03\downarrow} $&$ 5.63\pm0.16 $&$ -7\pm4 $&  & & &${57.9\pm2.1\downarrow} $& ${-1.2\pm0.7\downarrow}$& &0.76\\
 \multirow{-2}{*}{59206.66} &8.60& 1.95 &\multirow{-2}{*}{16926} &$48\pm9 $&$ 24.3\pm3.2 $&$ -4.90\pm1.51$&$- $&\multirow{-2}{*}{$-$}  &\multirow{-2}{*}{$-$} &$30\pm5 $&$ -1.6\pm0.6 $&\multirow{-2}{*}{$-$}&0.94\\
\midrule
&2.25 & 0.24 &  & $0.43\pm0.02 $&$ 6.26\pm0.24 $&$ 112\pm11 $&  & & &$42.1\pm2.0 $& $13.5\pm1.5$& &0.96\\
 \multirow{-2}{*}{59213.73} &8.60& 1.95 &\multirow{-2}{*}{5863} &$60\pm17 $&$ 29\pm6 $&$ -15.69\pm7.41$&$- $&\multirow{-2}{*}{$4.04\pm0.22$}  &\multirow{-2}{*}{$3.96\pm0.21$} &$28\pm7 $&$ -4.9\pm2.6 $&\multirow{-2}{*}{$9.9\pm1.9$}&0.88\\
\midrule
&{\color{red}2.25} & {\color{red}0.24} &  & ${\color{red}0.45\pm0.02} $&$ {\color{red}4.67\pm0.12} $&$ {\color{red}3.8\pm2.3} $&  & & &${\color{red}57.3\pm1.8} $& ${\color{red}0.6\pm0.4}$& &{\color{red}0.95}\\
\multirow{-2}{*}{\color{red}59221.63} &{\color{red}8.60}&{\color{red} 1.95} &\multirow{-2}{*}{\color{red}17979} &${\color{red}81\pm15} $&${\color{red} 14.8\pm1.9} $&$ {\color{red}-1.15\pm0.35}$&${\color{red}-} $&\multirow{-2}{*}{\color{red}$3.82\pm0.12$}  &\multirow{-2}{*}{\color{red}$4.20\pm0.14$} &${\color{red}64\pm10} $&$ {\color{red}-0.81\pm0.28} $&\multirow{-2}{*}{\color{red}$8.1\pm0.9$}&{\color{red}0.68}\\
\midrule
&2.25 & 0.24 &  & $0.37\pm0.01 $&$ 4.50\pm0.10 $&$ 71\pm4 $&  & & &$54.6\pm1.4 $& $11.1\pm0.7$& &0.92\\
 \multirow{-2}{*}{59223.61} &8.60& 1.95 &\multirow{-2}{*}{13379} &$51\pm11 $&$ 22.7\pm3.2 $&$ -6.08\pm2.04$&$66400\pm9700 $&\multirow{-2}{*}{$4.13\pm0.16$}  &\multirow{-2}{*}{$3.88\pm0.14$} &$33\pm6 $&$ -2.2\pm0.8 $&\multirow{-2}{*}{$12.3\pm1.7$}&0.90\\
\midrule
&{\color{blue}2.25} &{\color{blue} 0.24} &  & {\color{blue}$0.36\pm0.01 $}&{\color{blue}$ 4.30\pm0.10 $}&$ {\color{blue}64\pm4} $&  & & &{\color{blue}$55.6\pm1.7 $}& {\color{blue}$10.3\pm0.8$}& &{\color{blue}0.92}\\
 \multirow{-2}{*}{\color{blue}59224.61} &{\color{blue}8.60}& {\color{blue}1.95} &\multirow{-2}{*}{\color{blue}16235} &{\color{blue}$170\pm40 $}&{\color{blue}$ 12.6\pm2.0 $}&{\color{blue}$ 0.03\pm0.01$}&{\color{blue}$47200\pm1900 $}&\multirow{-2}{*}{\color{blue}$4.27\pm0.14$}  &\multirow{-2}{*}{\color{blue}$3.76\pm0.11$} &{\color{blue}$108\pm21 $}&{\color{blue}$ 0.038\pm0.019 $}&\multirow{-2}{*}{\color{blue}$14.7\pm1.9$}&{\color{blue}0.74}\\
\midrule
&{\color{blue}2.25} &{\color{blue} 0.24} &  & {\color{blue}$0.40\pm0.01 $}&{\color{blue}$ 4.84\pm0.12 $}&{\color{blue}$ -78\pm5 $}&  & & &{\color{blue}$52.4\pm1.5 $}& ${\color{blue}-11.7\pm0.9}$& &{\color{blue}0.95}\\
\multirow{-2}{*}{\color{blue}59225.57} &{\color{blue}8.60}& {\color{blue}1.95} &\multirow{-2}{*}{\color{blue}19302} &{\color{blue}$170\pm40 $}&{\color{blue}$ 13.5\pm2.0 $}&{\color{blue}$ -0.39\pm0.14$}&${\color{blue}- }$&\multirow{-2}{*}{\color{blue}$4.20\pm0.14$}  &\multirow{-2}{*}{\color{blue}$3.82\pm0.11$} &{\color{blue}$102\pm19 $}&{\color{blue}$ -0.43\pm0.17$}&\multirow{-2}{*}{\color{blue}$12.3\pm1.5$}&{\color{blue}0.91}\\
\midrule
&{\color{blue}2.25} &{\color{blue} 0.24} &  &{\color{blue} $0.42\pm0.01 $}&{\color{blue}$ 4.37\pm0.09 $}&{\color{blue}$ -6.4\pm2.3 $}&  & & &{\color{blue}$59.4\pm1.5 $}& ${\color{blue}-1.1\pm0.4}$& &{\color{blue}0.96}\\
\multirow{-2}{*}{\color{blue}59227.57} &{\color{blue}8.60}& {\color{blue}1.95} &\multirow{-2}{*}{\color{blue}15814} &{\color{blue}$150\pm40 $}&{\color{blue}$ 23\pm4 $}&{\color{blue}$ -1.90\pm0.85$}&{\color{blue}$- $}&\multirow{-2}{*}{\color{blue}$4.41\pm0.18$}  &\multirow{-2}{*}{\color{blue}$3.66\pm0.12$} &{\color{blue}$56\pm13 $}&{\color{blue}$ -1.2\pm0.6 $}&\multirow{-2}{*}{\color{blue}$13.7\pm2.1$}&{\color{blue}0.83}\\
\midrule
&2.25 & 0.24 &  & $0.32\pm0.01 $&$ 4.16\pm0.07 $&$ -12.3\pm2.3 $&  & & &$55.0\pm1.1 $& $-1.9\pm0.4$& &0.94\\
 \multirow{-2}{*}{59231.55} &8.60& 1.95 &\multirow{-2}{*}{19722} &$89\pm17 $&$ 15.4\pm2.0 $&$ 0.73\pm0.22$&$31900\pm2500 $&\multirow{-2}{*}{$4.08\pm0.12$}  &\multirow{-2}{*}{$3.92\pm0.11$} &$65\pm10 $&$ 0.51\pm0.18 $&\multirow{-2}{*}{$13.6\pm1.3$}&0.79\\
\midrule
&{\color{blue}2.25} &{\color{blue} 0.24} &  & {\color{blue}$0.36\pm0.01 $}&{\color{blue}$ 3.88\pm0.07 $}&{\color{blue}$ -0.7\pm1.9 $}&  & & &{\color{blue}$62.5\pm1.3 $}& {\color{blue}$-0.12\pm0.33$}& &{\color{blue}0.94}\\
\multirow{-2}{*}{\color{blue}59232.56} &{\color{blue}8.60}& {\color{blue}1.95} &\multirow{-2}{*}{\color{blue}17257} &${\color{blue}230\pm70} $&$ {\color{blue}28\pm6} $&$ {\color{blue}-1.09\pm0.57}$&{\color{blue}$28600\pm3200 $}&\multirow{-2}{*}{\color{blue}$4.87\pm0.21$}  &\multirow{-2}{*}{\color{blue}$3.39\pm0.10$} &{\color{blue}$58\pm16 $}&{\color{blue}$ -0.7\pm0.4 $}&\multirow{-2}{*}{\color{blue}$23\pm4$}&{\color{blue}0.95}\\
\midrule
&2.25 & 0.97 &  & ${ 0.60\pm0.02\downarrow} $&$ { 3.04\pm0.14\downarrow} $&$ { -8\pm5\downarrow} $&  & & &${ 102\pm5\downarrow} $& ${ -2.4\pm1.3\downarrow}$& &0.64\\
 \multirow{-2}{*}{59243.58} &8.60& 1.95 &\multirow{-2}{*}{16986} &$100\pm20 $&$ 14.5\pm2.0 $&$ -2.06\pm0.68$&$24500\pm2600 $&\multirow{-2}{*}{$-$}  &\multirow{-2}{*}{$-$} &$73\pm13 $&$ -1.6\pm0.6 $&\multirow{-2}{*}{$-$}&0.99\\
\midrule
&2.25 & 0.97 &  & ${ 0.74\pm0.03\downarrow} $&$ { 3.11\pm0.10\downarrow} $&$ { 19.1\pm3.5\downarrow} $&  & & &${ 110\pm5\downarrow} $& ${ 6.0\pm1.1\downarrow}$& &0.78\\
 \multirow{-2}{*}{59293.46} &8.60& 1.95 &\multirow{-2}{*}{15844} &$94\pm17 $&$ 10.6\pm1.3 $&$ -1.92\pm0.56$&$4500\pm60 $&\multirow{-2}{*}{$-$}  &\multirow{-2}{*}{$-$} &$96\pm15 $&$ -2.0\pm0.7 $&\multirow{-2}{*}{$-$}&1.00\\
\midrule
&{\color{red}2.25} &{\color{red} 0.24} &  &{\color{red} $0.44\pm0.01 $}&{\color{red}$ 2.87\pm0.07 $}&{\color{red}$ 25.1\pm3.4 $}&  & & &{\color{red}$93.1\pm2.7 $}& {\color{red}$6.7\pm0.9$}& &{\color{red}0.97}\\
\multirow{-2}{*}{\color{red}59295.45} &{\color{red}8.60}&{\color{red} 1.95} &\multirow{-2}{*}{\color{red}14040} &{\color{red}$58\pm9 $}&{\color{red}$ 10.0\pm1.1 $}&{\color{red}$ -1.92\pm0.49$}&{\color{red}$5230\pm120 $}&\multirow{-2}{*}{\color{red}$3.77\pm0.11$}  &\multirow{-2}{*}{\color{red}$4.26\pm0.14$} &{\color{red}$81\pm11 $}&{\color{red}$ -1.7\pm0.5 $}&\multirow{-2}{*}{\color{red}$7.8\pm0.8$}&{\color{red}0.97}\\
\midrule
&2.25 & 0.24 &  & $0.44\pm0.01 $&$ 2.57\pm0.04 $&$ -27.7\pm2.4 $&  & & &$104.4\pm2.2 $& $-8.2\pm0.7$& &0.95\\
 \multirow{-2}{*}{59308.35} &8.60& 1.95 &\multirow{-2}{*}{17287} &$139\pm25 $&$ 8.8\pm1.1 $&$ 0.89\pm0.26$&$2886\pm28 $&\multirow{-2}{*}{$4.10\pm0.11$}  &\multirow{-2}{*}{$3.91\pm0.10$} &$141\pm21 $&$ 1.4\pm0.4 $&\multirow{-2}{*}{$10.0\pm0.9$}&0.85\\
\midrule
&{\color{blue}2.25} &{\color{blue} 0.24} &  & {\color{blue}$0.45\pm0.01 $}&{\color{blue}$ 1.67\pm0.03 $}&{\color{blue}$ -29.1\pm2.6 $}&  & & &{\color{blue}$163\pm4 $}& {\color{blue}$-13.5\pm1.3$}& &{\color{blue}0.95}\\
\multirow{-2}{*}{\color{blue}59749.17} &{\color{blue}8.60}& {\color{blue}1.95} &\multirow{-2}{*}{\color{blue}15243} &{\color{blue}$420\pm90 $}&{\color{blue} $7.0\pm1.1 $}&{\color{blue}$ 1.02\pm0.37$}&{\color{blue}-}&\multirow{-2}{*}{\color{blue}$4.72\pm0.14$}  &\multirow{-2}{*}{\color{blue}$3.47\pm0.07$} &{\color{blue}$310\pm60\downarrow $}&{\color{blue}$ 3.4\pm1.4 $}&\multirow{-2}{*}{\color{blue}$16.1\pm1.9$}&{\color{blue}0.99}\\
\midrule
&{\color{blue}2.25} &{\color{blue} 0.24} &  & {\color{blue}$0.54\pm0.01 $}&{\color{blue}$ 1.88\pm0.03 $}&{\color{blue}$ -2.8\pm1.7 $}&  & & &{\color{blue}$158.2\pm3.2 $} & {\color{blue}$-1.3\pm0.8$}& &{\color{blue}0.97}\\
\multirow{-2}{*}{\color{blue}59755.14} &{\color{blue}8.60}& {\color{blue}1.95} &\multirow{-2}{*}{\color{blue}20895} &{\color{blue}$160\pm26 $} &{\color{blue}$ 7.7\pm0.9 $}&{\color{blue}$ -0.15\pm0.04$}&{\color{blue}$1148\pm25 $}&\multirow{-2}{*}{\color{blue}$4.17\pm0.10$}  &\multirow{-2}{*}{\color{blue}$3.84\pm0.09$} &{\color{blue}$173\pm24 $}&{\color{blue}$ -0.28\pm0.08 $}&\multirow{-2}{*}{\color{blue}$8.6\pm0.8$}&{\color{blue}1.06}\\
\midrule
&{\color{blue}2.25} &{\color{blue} 0.24} &  &{\color{blue} $0.44\pm0.01 $}&{\color{blue}$ 1.81\pm0.03 $}&{\color{blue}$ 12.0\pm1.0 $ }&  & & &{\color{blue}$149.0\pm2.6 $}& {\color{blue}$5.1\pm0.4$}& &{\color{blue}0.97}\\
\multirow{-2}{*}{\color{blue}59791.04} &{\color{blue}8.60}& {\color{blue}1.95 }&\multirow{-2}{*}{\color{blue}15333} &{\color{blue}$740\pm230 $}&{\color{blue}$ 14.8\pm3.2 $}&{\color{blue}$ -0.18\pm0.09$}&{\color{blue}$- $}&\multirow{-2}{*}{\color{blue}$5.36\pm0.19$}  &\multirow{-2}{*}{\color{blue}$3.19\pm0.07$} &{\color{blue}$190\pm50$} &{\color{blue}$ -0.38\pm0.22 $}&\multirow{-2}{*}{\color{blue}$28\pm4$}&{\color{blue}0.95}\\
\midrule
&{\color{blue}2.25} &{\color{blue} 0.24} &  & {\color{blue}$0.39\pm0.01 $}&{\color{blue}$ 1.95\pm0.03 $}&{\color{blue}$ 1.7\pm1.3 $}&  & & &{\color{blue}$129.4\pm2.2 $}& {\color{blue}$0.6\pm0.5$}& &{\color{blue}0.96}\\
\multirow{-2}{*}{\color{blue}59812.98} &{\color{blue}8.60}& {\color{blue}1.95} &\multirow{-2}{*}{\color{blue}17618} &{\color{blue}$146\pm25 $}&{\color{blue}$ 8.1\pm1.0 $}&{\color{blue}$ 0.73\pm0.20$}&{\color{blue}$1970\pm40 $}&\multirow{-2}{*}{\color{blue}$4.28\pm0.11$}  &\multirow{-2}{*}{\color{blue}$3.75\pm0.08$} &{\color{blue}$158\pm23 $}&{\color{blue}$ 1.2\pm0.4 $}&\multirow{-2}{*}{\color{blue}$13.1\pm1.2$}&{\color{blue}1.05}\\
\midrule
&{\color{blue}2.25} &{\color{blue} 0.24} &  &{\color{blue} $0.39\pm0.01 $}&{\color{blue}$ 2.07\pm0.03 $}&{\color{blue}$ 0.1\pm2.3 $}&  & & &${\color{blue}122.2\pm2.4 }$& ${\color{blue}0.0\pm0.8}$& &{\color{blue}0.96}\\
\multirow{-2}{*}{\color{blue}59832.92} &{\color{blue}8.60}& {\color{blue}1.95} &\multirow{-2}{*}{\color{blue}15904} &{\color{blue}$84\pm16 $}&{\color{blue}$ 12.1\pm1.6 $}&{\color{blue}$ 1.48\pm0.46$}&{\color{blue}$- $}&\multirow{-2}{*}{\color{blue}$4.38\pm0.14$}  &\multirow{-2}{*}{\color{blue}$3.68\pm0.10$} &{\color{blue}$80\pm13 $}&{\color{blue}$ 1.3\pm0.4 $}&\multirow{-2}{*}{\color{blue}$14.1\pm1.7$}&{\color{blue}0.98}\\
\midrule
&{\color{red}2.25} &{\color{red} 0.24} &  & {\color{red}$0.53\pm0.02 $}&{\color{red}$ 2.61\pm0.06 $}&{\color{red}$ -4.4\pm1.5 $}&  & & &{\color{red}$112.1\pm3.0 $}& {\color{red}$-1.4\pm0.5$}& &{\color{red}0.99}\\
\multirow{-2}{*}{\color{red}59852.85} &{\color{red}8.60}&{\color{red} 1.95} &\multirow{-2}{*}{\color{red}21466} &{\color{red}$54\pm7 $}&{\color{red}$ 9.6\pm0.8 $}&{\color{red}$ 0.44\pm0.09$}&{\color{red}$- $}&\multirow{-2}{*}{\color{red}$3.74\pm0.09$}  &\multirow{-2}{*}{\color{red}$4.30\pm0.12$} &{\color{red}$81\pm9 $}&{\color{red}$ 0.39\pm0.09 $}&\multirow{-2}{*}{\color{red}$6.3\pm0.5$}&{\color{red}1.16}\\
\midrule
&{\color{red}2.25} &{\color{red} 0.24} &  & {\color{red}$0.48\pm0.01 $}&{\color{red}$ 2.81\pm0.06 $}&{\color{red}$ -42.0\pm2.3 $}&  & & &{\color{red}$99.2\pm2.4 $}& {\color{red}$-11.9\pm0.7$}& &{\color{red}0.93}\\
\multirow{-2}{*}{\color{red}59861.83} &{\color{red}8.60}& {\color{red}1.95} &\multirow{-2}{*}{\color{red}14311} &{\color{red}$41\pm6 $}&{\color{red}$ 9.3\pm0.9 $}&{\color{red}$ -1.27\pm0.29$}&{\color{red}$- $}&\multirow{-2}{*}{\color{red}$3.58\pm0.10$}  &\multirow{-2}{*}{\color{red}$4.52\pm0.16$} &{\color{red}$72\pm9 $}&{\color{red}$ -1.01\pm0.26 $}&\multirow{-2}{*}{\color{red}$6.1\pm0.5$}&{\color{red}0.90}\\
\bottomrule
\end{tabular}}
Note: Upper limit results are labeled with $\downarrow$. Results that follow standard ISS theory and ``anomalous scattering'' are labeled with {\color{blue}blue} and {\color{red}red} characters, respectively.
\label{tab:ourresults}
\end{table*}

\subsection{Dynamic spectra}
\label{sec:dymspec}
We used dynamic spectrum plots to display the DISS phenomena observed on PSR~B0740$-$28. As shown in two pairs of representative plots in Figure~\ref{fig:opra}(b), the dynamic spectrum plots for the 2.25~$\&$~8.60~GHz observations are presented with different color styles. $S$ is the intensity at each pixel of the dynamic spectrum plot, while $S_{\rm max}$ is the maximum intensity within the observation (see supplementary material \href{run:.supplement.pdf}{Figure S1 online} for all dynamic spectrum plots for our observations). It is evident that the typical size of the scintillation patterns (also called `scintles') at 2.25~GHz was much smaller than that at 8.60~GHz. During the observational period, not only did the size and shape of the scintles change, but their regularities also varied from epochs to epochs. Scintles were sometimes randomly distributed and sometimes regularly drifting. Taking Figure~\ref{fig:opra}(b) as an example, it can be seen that the scintles were vertical with the time axis at 8.60~GHz on MJD~59225.57, while the scintles drifted from the top-left to the bottom-right on MJD~59295.45 at the same $\nu$. In accordance with most previous research \citep{Hew80, Ric90}, the size of the scintles will subsequently be quantified with $\Delta\nu_{\rm d}$ and $\Delta\tau_{\rm d}$. Meanwhile, the frequency drift rate ($d\nu/dt$ or $dt/d\nu$) is used to characterize their drifting features.

\subsection{Two-dimensional autocovariance function}
The two-dimensional (2D) autocovariance function (ACF) is a widely used method for quantifying the average characteristics of scintles detected in the dynamic spectrum plots. Based on the intensity $S(\nu,t)$ for the pixel $(\nu,t)$ of the dynamic spectrum plot and the corresponding average intensity $\overline{S(\nu,t)}$, the 2D-ACF is normally computed with the following equation:
\begin{equation}
F(\Delta\nu, \Delta\tau)=\sum_{\nu}\sum_{t} {\Delta}S(\nu,t) {\Delta}S(\nu+\Delta\nu,t+\Delta\tau)
\end{equation}
where ${\Delta}S(\nu,t)=S(\nu,t)-\overline{S(\nu,t)}$ represents the intensity fluctuation for each pixel. To clearly show related fluctuations on the plots, the 2D-ACF results are further normalized by $\rho(\Delta\nu, \Delta\tau) = F(\Delta\nu, \Delta\tau)/F(0, 0)$. 

In order to more accurately fit the 2D-ACF results and to more precisely quantify the average size of the scintles, researchers had tested the 2D elliptical Gaussian function \citep{grl94, brg99a} as well as the Lorentzian functions \citep{gbr98, tdc24}. However, it was found that the fitting results for $\Delta\nu_{\rm d}$ were sometimes affected by the skewness of the 2D-ACF, which was caused by the refractive shift effects \citep{rcn14}. Consequently, the 2D-Exponential form function was used in several recent studies \citep{rcn14, rch19, rc23}. This function can be expressed as:
\begin{equation}
\begin{aligned}
\rho(\Delta\nu, \Delta\tau)&=Ae^{-(|\frac{\Delta\tau-\frac{dt}{d\nu}\Delta\nu}{\Delta\tau_{\rm d}}|^{\rm \frac{3\alpha_{\rm 0}}{2}} +|\frac{\Delta\nu}{\Delta\nu_{\rm d}/ln2}|^{\frac{3}{2}})^{\rm \frac{2}{3}}}\\&*\Lambda(\Delta\tau,T)\Lambda(\Delta\nu,BW)
\end{aligned}
\end{equation}
and $\rho(0, 0)=A+w=1$, where $A$ and $w$ are the parameters used to smooth the white noise peak located at (0, 0). The triangle functions, $\Lambda(\Delta\tau, T)=1 -\Delta\tau/T$ and $\Lambda(\Delta\nu, BW)=1 -\Delta\nu/BW$, are employed to reduce the influence of finite observation length $T$ and bandwidth $BW$, respectively. Meanwhile, $\alpha_{\rm 0}$ is set to be 5/3 \citep{rc23}. As previous research had revealed, the 2D-Exponential form function provided a superior fit to the other two functions. Consequently, the following analysis is based on the results obtained with this function.

For the convenience of fitting our 2D-ACF results, we called the open-source software \texttt{scintools} \citep{rc23,rcb20}\footnote{\label{foot:scin}\url{https://github.com/danielreardon/scintools}}. Figure~\ref{fig:opra}(a) presents plots of the normalized 2D-ACFs along with the best-fitting results corresponding to the dynamic spectra shown in Figure~\ref{fig:opra}(b) (see \href{run:.supplement.pdf}{Figure S2 online} for similar 2D-ACF plots for all of our observations). As previously mentioned, their white noise peaks at (0, 0) have been smoothed by parameters $A$ and $w$ from the 2D-Exponential form function.

In Table~\ref{tab:ourresults}, columns 5$-$7 are the best-fitting results for $\Delta\nu_{\rm d}$, $\Delta\tau_{\rm d}$, and $dt/d\nu$ from our observations. These parameters were obtained by fitting the global 2D-ACF results with the 2D-Exponential form function. As in most previous studies, we considered two primary sources of error for our fitting results of related parameters and added them in quadrature. Besides the fitting errors, we also considered the fractional errors, $\sigma$, caused by the finite $T$ and $BW$. The $\sigma$ was obtained with:
\begin{equation}
\sigma=\sqrt{(\frac{\Delta\tau_{\rm d}}{T})(\frac{\Delta\nu_{\rm d}}{BW})f^{\rm -1}}
\end{equation}
where $f$ is the ratio between the area of bright scintles and the area of a single feature in the dynamic spectrum plots \citep{bgr99}. In our calculation, $f$ is set to 0.5. In addition, the fitted $\Delta\nu_{\rm d}$ and $\Delta\tau_{\rm d}$ were corrected for smearing due to finite instrumental resolutions in both frequency and time using a quadrature subtraction scheme \citep{bgr99}. It is evident that the typical $\Delta\nu_{\rm d}$ of PSR~B0740$-$28 at 2.25~GHz was less than 1.0~MHz, based on the results presented in Table~\ref{tab:ourresults}. Due to the insufficient frequency resolution ($\Delta\nu_{\rm 0}=0.97$~MHz) used in our 2.25~GHz observations on MJDs~59168.83, 59206.66, 59243.58, and 59293.46, we are only able to determine the upper limits of $\Delta\nu_{\rm d}$ and related results (labeled with $\downarrow$).

\subsection{Secondary spectra}
\label{sec:secondaryspec}
The Fourier power spectrum of the pulsar's dynamic spectrum - the secondary spectrum ($\mathcal{S}$) - sometimes exhibits a parabolic distribution of power. This phenomenon can be attributed to the existence of discrete scattering screens located at the line of sight to the pulsar \citep{cw86, smc01, wms04, crs06}. In order to remove the frequency dependence of the arc curvature, we used the \texttt{scintools}$^{\ref{foot:scin}}$ to uniformly re-sample the dynamic spectrum in wavelength, thereby obtaining $S(t, \lambda)$. Meanwhile, the Hamming window function is applied to the outer 10\% of each dynamic spectrum to reduce the side lobe response. Before using the 2D Fourier transform to calculate the secondary spectrum, \texttt{scintools} subtracts the mean flux from $S(t, \lambda)$ and pre-whitens it. In order to improve the visibility of arc structures in the corresponding plots, a logarithmic secondary spectrum is calculated according to the following expression:
\begin{equation}
\mathcal{S}(f_{\rm t},f_{\lambda})=10log_{10}(|\tilde{S}(t, \lambda)|^{\rm 2})
\label{equ:ss}
\end{equation}
where tilde denotes the 2D Fourier transform, and $\tilde{S}(t, \lambda)$ is the mean-subtracted and windowed dynamic spectrum with wavelength $\lambda$ \citep{fcm14}.

In Figure~\ref{fig:opra}(c), we present two sample pairs of secondary spectrum plots corresponding to the dynamic spectra shown in Figure~\ref{fig:opra}(b). In order to improve the visibility of the arc structures, we zoomed in on the regions where arcs can be seen in these plots (see \href{run:.supplement.pdf}{Figure S3 online} for all secondary spectra). Although we could not discern arc structures in any of the secondary spectrum plots for our 2.25~GHz observations of PSR~B0740$-$28, arc structures were discernible in similar plots of 8.60~GHz observations between MJDs~59223.61 and 59812.98 (except for MJDs~59225.57, 59227.57, 59749.17, and 59791.04). The absence of discernible arcs in the secondary spectrum plots for our 2.25~GHz observations is likely attributable to the small scintles, which are insufficient to contribute enough power to the Fourier power spectra. When the arc is distinctly observable as the red-dashed parabolic curve marked in the rightmost panel of Figure~\ref{fig:opra}(c), the degree of arc curvature ($\eta$) is further fitted using \texttt{scintools} with the equation $f_{\rm \lambda}={\eta}f_{\rm t}^{\rm 2}$. In this equation, $f_{\rm t}$ and $f_{\rm \lambda}$ are the Fourier conjugates of the time $t$ and wavelength $\lambda$ axes of the dynamic spectrum, respectively. The manner in which \texttt{scintools} fits $\eta$ is briefly introduced in \ref{arc}. Additionally, we have included related plots about $\eta$ fitting (such as normalized secondary spectra, the Doppler profiles, and the results of the arc curvature fitting) in the online supplementary materials (see \href{run:.supplement.pdf}{Figure S4 online}). The best-fitting values of $\eta$ are presented in column 8 of Table~\ref{tab:ourresults}.

\section{Discussions}
\label{sec:discuss}
\subsection{Comparison with previous observation results}
The comparison of the historical results summarized in Table~\ref{tab:preresults} with our results shown in Table~\ref{tab:ourresults} reveals significant advancements in the ISS observation of PSR~B0740$-$28. The $\Delta\nu_{\rm d}$ was too narrow to be observed at 0.66~GHz with the Parkes telescope, so only $\Delta\tau_{\rm d}$ was fitted successfully \citep{jnk98}. The values of $\Delta\nu_{\rm d}$ at 2.25~GHz were reported to be $1.1\pm0.1$, $1.1\pm0.1$, and $0.6\pm0.1$~MHz by the Jiamusi telescope with frequency resolutions, $\Delta\nu_{\rm 0}$, of 0.58, 1.17, and 0.58~MHz, respectively, using 2D-Gaussian fitting \citep{whh18}. In comparison, the values of $\Delta\nu_{\rm d}$ obtained by our observations at 2.25~GHz range from 0.32 to 0.54~MHz, except for four results affected by insufficient $\Delta\nu_{\rm 0}$. This finding suggested that the values of $\Delta\nu_{\rm d}$ obtained by the Jiamusi telescope at 2.25~GHz were also affected by insufficient $\Delta\nu_{\rm 0}$. For the ISS observations of this pulsar at 4.75~GHz and 10.55~GHz with the Effelsberg telescope, as more attention was paid to the flux density modulation effects of weak ISS, only $\Delta\tau_{\rm d}$ was obtained \citep{mss96}. The researchers concluded that the critical frequency between the strong and weak ISS should be at least 5.0~GHz. Our observations indicated that the ISS of PSR~B0740$-$28 remained within the strong scintillation region at 8.60~GHz. This conclusion is supported by both the scintillation strength $u = \sqrt{2\nu/\Delta\nu_{\rm d}} >> 1.0$ \citep{Ric90} and the modulation index $m=\sigma_{\rm S}/\overline{S(\nu,t)}\approx1$, where $\sigma_{\rm S}$ is the standard deviation of flux intensities. The results for $m$ are also presented in the last column of Table.~\ref{tab:ourresults}.

\subsection{The fluctuation of electron density}
For the ISM with a scattering spectral index of $\alpha$, the observed $\Delta\nu_{\rm d}$ and $\Delta\tau_{\rm d}$ will theoretically change with $\nu$ following the power law, which can be expressed as $\Delta\nu_{\rm d} \propto \nu^{\rm \alpha}$ and $\Delta\tau_{\rm d} \propto \nu^{(\rm \alpha-2)/2}$, respectively \citep{Ric77}. For a dual-frequency ISS observation, they can be further transformed into:
\begin{equation}
\Delta\nu_{\rm d,1}/\Delta\nu_{\rm d,2}=({\nu_{\rm 1}}/{\nu_{\rm 2}})^{\rm \alpha}   
\label{equa}
\end{equation}
and
\begin{equation}
(\Delta\tau_{\rm d,1}/\Delta\tau_{\rm d,2})^2=({\nu_{\rm 1}}/{\nu_{\rm 2}})^{\rm \alpha-2}
\label{equb}
\end{equation}
where [$\Delta\nu_{\rm d,1}$, $\Delta\tau_{\rm d,1}$] and [$\Delta\nu_{\rm d,2}$, $\Delta\tau_{\rm d,2}$] are the [$\Delta\nu_{\rm d}$, $\Delta\tau_{\rm d}$] obtained at frequencies $\nu_{\rm 1}$ and $\nu_{\rm 2}$, respectively. Combining Eq.~\ref{equa} and Eq.~\ref{equb}, we can derive the following equation:
\begin{equation}
\frac{\Delta\nu_{\rm d,1}}{\Delta\nu_{\rm d,2}}+(\frac{\Delta\tau_{\rm d,1}}{\Delta\tau_{\rm d,2}})^{\rm 2}=(1+(\frac{\nu_{\rm 1}}{\nu_{\rm 2}})^{-2})(\frac{\nu_{\rm 1}}{\nu_{\rm 2}})^{\rm \alpha}
\label{equ:alpha}
\end{equation}
Eq.~\ref{equ:alpha} can be used to jointly calculate the values of $\alpha$ with both $\Delta\nu_{\rm d}$ and $\Delta\tau_{\rm d}$. In column 9 of Table.~\ref{tab:ourresults}, we present the values of $\alpha$ obtained with Eq.~\ref{equ:alpha} based on our simultaneous 2.25~$\&$~8.60~GHz ISS observation results presented in columns 5$-$6 of this table. The uncertainties of $\alpha$ were estimated by the Python module \texttt{uncertainties} \footnote{\label{foot:uncer}\url{http://pythonhosted.org/uncertainties/}}\citep{uncerLE}, which calculates uncertainties using the linear error propagation theory. The uncertainties of the parameters in Eq.~\ref{equ:alpha} are independent of each other, so the \texttt{uncertainties} module added them in quadrature. With the exception of four epochs of observations affected by insufficient frequency resolution at 2.25~GHz (on MJDs~59168.83, 59206.66, 59243.58, and 59293.46) and another four epochs (on MJDs~59213.73, 59223.61, 59231.55, and 59308.35) with nominal values of $\alpha \approx 4.0$, the other thirteen epochs of observations with $\alpha < 4.0$ and $\alpha>4.0$ in their error ranges are labeled with red and blue characters in Table.~\ref{tab:ourresults}, respectively. It is evident that four epochs of our observations yield $\alpha < 4.0$. For interstellar plasma with a purely Kolmogorov spectrum, the expectation was that $\alpha = 4.4$. Other turbulence models may yield different values of $\alpha$, but the lowest value for a thin screen model is $\alpha = 4.0$ (for the so-called critical turbulence model with $\beta = 4$) \citep{rnb86,lkk15,kmj19}. Previous studies have also reported $\alpha < 4.0$ in multi-frequency observations about DISS \citep{jnk98, wmj05, whh18} and interstellar scattering, which were classified as anomalous scattering \citep{lkm01, lkk15, kmj19}. However, It is notable that the majority of the data used in the above research, particularly the DISS observations, were not obtained simultaneously. The time intervals between observations were sometimes as long as several years, suggesting that the interstellar scattering and scintillation may not be caused by the same ISM clumps. Therefore, our simultaneous dual-frequency observations provide stronger support for $\alpha < 4.0$.

According to the ISS theory, the dynamic spectrum will be independent of frequency if $\beta > 4.0$ \citep{gn85,kmj19}. Due to the evident discrepancies between the dynamic spectra of the 2.25~GHz and 8.60~GHz observations presented in Figure~\ref{fig:opra}b, we employed the relation $\alpha= 2\beta/(\beta-2)$ to obtain related $\beta$ values \citep{Ric77} and present them in column 10 of Table~\ref{tab:ourresults}. The uncertainties of $\beta$ were also estimated with the Python module \texttt{uncertainties}$^{\ref{foot:uncer}}$. Although this relationship is suitable for $\beta < 4.0$ only, we obtained $\beta > 4.0$ when anomalous scattering occurs ($\alpha < 4.0$). Similar inexplicable results were also found in some earlier observations \citep{lkm01,wmj05,ldk13}. Theoretical analysis indicates that anisotropic irregularities may lead to variations in the broadening functions for different frequencies and cause anomalous scattering \citep{cl01}.

The pulse scatter broadening timescale ($t_{\rm d}$) is an important parameter used to describe the interstellar scattering effect. In column 13 of Table.~\ref{tab:ourresults}, we present the estimated broadening timescale at 1~GHz, $t_{\rm d, 1~GHz}$, which is obtained with the relationship $2\pi t_{\rm d}\Delta\nu_{\rm d}=1$ \citep{cr98} and $t_{\rm d}\propto \nu^{-\alpha}$ (a simplified expression of Eq.~\ref{equa}). The average $t_{\rm d, 1~GHz}$ of our observations is about $13\pm5~{\mu}$s, which is consistent with the predicted result $t_{\rm d, 1~GHz}=12\pm6~{\mu}$s using the previous interstellar scattering measurements at 160~MHz ($t_{\rm d, 160~MHz}=24.5\pm2.8$~ms) \citep{abs86} and the average $\alpha$ of our observations ($\alpha=4.23\pm0.33$). The comparison of PSR~B0740$-$28 with other pulsars in the PSRCAT that have similar dispersion measures ($\rm DM=73.78~cm^{\rm -3}pc$) \citep{mht05,pkj13} reveals that only a few pulsars show larger $t_{\rm d, 1~GHz}$ than that of PSR~B0740$-$28. Furthermore, the interstellar scattering caused by the fluctuations in the Gum Nebula could explain the enhanced scattering of PSR~B0740$-$28 and other pulsars in this direction \citep{mr01}. Thus, the estimated scatter broadening timescales derived from our observations support the idea that the Gum Nebula enhanced the scattering of PSR~B0740$-$28.

\subsection{Location and motion of the scintillation screen}
\label{sec:screenfit}

\begin{figure*}[ht]
	\centering                                               
	{\begin{minipage}[t]{1\linewidth}
	\centering
	\includegraphics[height=5cm,width=17cm]{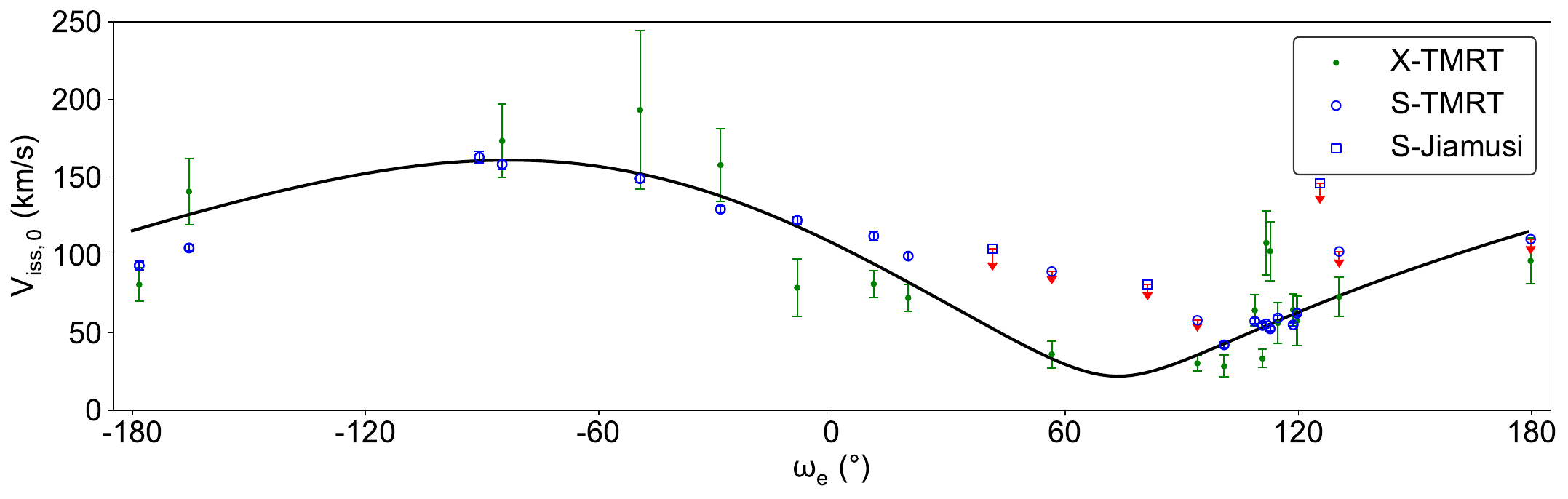}   
	\end{minipage}}
    \caption{$V_{\rm iss, 0}$ varies as a function of $\omega_{\rm e}$. The curve is fitted based on Eq.~\ref{visseclip2} using 2.25 and 8.60~GHz observation results, except for those marked with upper limit markers ($\downarrow$).}  
    \label{fig:scrfitplt}
    \end{figure*}

If we assume that the scintillation screen is located at a distance $L_{\rm o}$ from the observer and at a distance  $L_{\rm p}$ from the pulsar, then the distance of the pulsar from the observer is $D= L_{\rm o}+L_{\rm p}$. The observed ISS speed ($V_{\rm iss}$) can be obtained with the following equation:
\begin{equation}
    V_{\rm iss}=A_{\rm 0}\sqrt{\Delta\nu_{\rm d} D x}/(\nu \Delta\tau_{\rm d})=\sqrt{x}V_{\rm iss,0}
    \label{eq:vissbasic}
\end{equation}
where $x = L_{\rm o}/L_{\rm p}$ and the reduced ISS speed $V_{\rm iss,0} = A_{\rm 0}\sqrt{\Delta\nu_{\rm d} D}/(\nu\Delta\tau_{\rm d})$ is $V_{\rm iss}$ for the special case of $x = 1.0$. Normally, the model-dependent coefficient $A_{\rm 0}$ is $3.85\times10^{\rm 4}$~km/s when the units of $\Delta\nu_{\rm d}$, $D$, $\nu$, and $\Delta\tau_{\rm d}$ are MHz, kpc, GHz, and seconds, respectively \citep{Gup95}.

In principle, $\vec{V}_{\rm iss}$ is a vector combination of the proper-motion velocity of the pulsar ($\vec{V}_{\rm p, \bot}$), the orbital velocity of the Earth ($\vec{V}_{\rm earth, \bot}$), and the transverse velocity of the scintillation screen ($\vec{V}_{\rm scr, \bot}$), which can be expressed as:
\begin{equation}
\vec{V}_{\rm iss}=x\vec{V}_{\rm p, \bot}+\vec{V}_{\rm earth, \bot}-(1+x)\vec{V}_{\rm scr, \bot}
\label{eq:vissvec}
\end{equation}
where $\bot$ stands for the projection component that is vertical to the line of sight. For a solitary pulsar,  the $\vec{V}_{\rm p, \bot}$ and $\vec{V}_{\rm scr, \bot}$ can be treated as constants in observations conducted over several years \citep{cr98,xsl23}. In the barycentric ecliptic coordinate system, $\vec{V}_{\rm earth}$ can be decomposed into the components $\dot{Z}$, $\dot{X}$, and $\dot{Y}$. Meanwhile, $\vec{V}_{\rm p, \bot}$, $\vec{V}_{\rm scr, \bot}$, and $\vec{V}_{\rm iss}$ can be decomposed into [$V_{\rm p,\omega}$, $V_{\rm p,\delta}$], [$V_{\rm scr, \omega}$, $V_{\rm scr, \delta}$], and [$V_{\rm iss, \omega}$, $V_{\rm iss, \delta}$] by projecting along the ecliptic longitude ($\omega$) and the ecliptic latitude ($\delta$), respectively. For a solitary pulsar located at [$\omega_{\rm p}, \delta_{\rm p}$] in the barycentric ecliptic coordinate system, the components $V_{\rm iss, \omega}$ and $V_{\rm iss, \delta}$ can be obtained from the following equations
\begin{equation}
       {V}_{\rm iss,\delta}=xV_{\rm p,\delta}-(1+x)V_{\rm scr,\delta}-\dot{X}sin{\delta_{\rm p}}cos{\omega_{\rm p}}-\dot{Y}sin{\delta_{\rm p}}sin{\omega_{\rm p}}+\dot{Z}cos{\delta_{\rm p}}
\label{eq:vissvs}
\end{equation}
and 
\begin{equation}
{V}_{\rm iss,\omega}=xV_{\rm p,\omega}-(1+x)V_{\rm scr,\omega}-(\dot{X}sin{\omega_{\rm p}}-\dot{Y}cos{\omega_{\rm p}})  
\label{eq:vissvw}
\end{equation}
respectively \citep{bpl02}. The Earth's orbit is nearly a circle on the ecliptic plane, so we adopted the following equations: $\dot{Z}=0$, $\dot{X}=V_{\rm e} sin{\omega_{\rm e}}$, and $\dot{Y}=-V_{\rm e} cos{\omega_{\rm e}}$, where $V_{\rm e}$ and $\omega_{\rm e}$ are the Earth's revolution speed and the ecliptic longitude of the Earth. Based on the proper motion of the pulsar in the Ecliptic coordinate system [$\mu_\omega$, $\mu_\delta$] in mas/yr, [$V_{\rm p,\omega}$, $V_{\rm p,\delta}$] can be obtained with $4.74D[\mu_\omega, \mu_\delta]$ in km/s. Considering $V_{\rm iss}=\sqrt{x}V_{\rm iss,0}$ (see Eq.~\ref{eq:vissbasic}), we substitute these variables into Eq.~\ref{eq:vissvs} and Eq.~\ref{eq:vissvw} to obtain the following equations:
\begin{equation}
\begin{aligned}
    &xV_{\rm iss,0}^2=[4.74xD\mu_\omega-(1+x)V_{\rm scr,\omega}-
    V_{\rm e}cos(\omega_{\rm e}-\omega_{\rm p})]^2\\&
    +[4.74xD\mu_\delta-(1+x)V_{\rm scr,\delta}-
    V_{\rm e}sin{\delta_{\rm p}}sin(\omega_{\rm e}-\omega_{\rm p})]^2
\end{aligned}
\label{visseclip2}
\end{equation}
in which there are three free parameters ($x$, $V_{\rm scr, \delta}$, and $V_{\rm scr, \omega}$) that need to be fitted. Our observations covered a wide range of $\omega_{\rm e}$, which enabled us to directly fit these three parameters using the least-squares curve fitting method.

In Figure~\ref{fig:scrfitplt}, we present the curve that corresponds to the best-fitting result. The best-fitting results for our $V_{\rm iss, 0}$ presented in Table~\ref{tab:ourresults} (except for 4 epochs of 2.25~GHz observations affected by inadequate $\Delta\nu_{\rm 0}$ and the 8.60~GHz observation on MJD~59749.17 affected by limited effective bandwidth caused by serious RFIs) are $V_{\rm scr,\delta} = 17.3^{+3.2}_{-3.4} $~km/s, $V_{\rm scr,\omega} = -14.3^{+5.4}_{-5.5} $~km/s, and $x = 0.14^{+0.09}_{-0.06}$. In addition, three historical $V_{\rm iss, 0}$ results measured by the Jiamusi telescope \citep{whh18} are also presented and labeled with red downward arrows in Figure~\ref{fig:scrfitplt}. Although these seven $V_{\rm iss, 0}$ results are not incorporated into the curve fitting, they are in good agreement with our best-fitting curve. Based on our best-fitting $x$ and the distance of PSR~B0740$-$28, we obtain the distance of the screen ($L_{\rm o} = 245^{+69}_{-72}$~pc) and the speed of the screen ($|\vec{V}_{\rm scr, \bot}| = 22.4^{+4.1}_{-4.2}$~km/s), respectively. Coincidentally, the Gum Nebula is centered around the Galactic coordinate [$258^\circ$, $-2^\circ$] with a radius of $\sim$130~pc and a thickness of $\sim$~20~pc at a distance of $\sim$350~pc \citep{cs83, sko22}. This indicates that the distance from its front edge to the observer is $\sim$240~pc, which is consistent with the best-fitting distance of the scattering screen.

Based on the differential rotation of the Milky Way, $\Vec{V}_{\rm scr, \bot}$ ought to be the vector sum of the expanding velocity of the Gum Nebula ($\vec{V}_{\rm exp, \bot}$) and its relative velocity to the Sun ($\Vec{V}_{\rm r, \bot}$). Assuming that the Sun is located at a distance of $8.12\pm0.03$~kpc from the Galactic center with a rotation speed of $233.6\pm2.8$~km/s and a velocity gradient of about $-1.34\pm0.21~{\rm pc^{\rm -1} km/s}$ \citep{mua19}, we infer that $\Vec{V}_{\rm r, \bot}$ is about 9~km/s. The basic vector operations indicated that $\vec{V}_{\rm exp, \bot}$ was approximately 13~km/s, which was in good agreement with the $\vec{V}_{\rm exp, \bot}$ obtained from the Doppler shift measurements of the $H\alpha$, N~II, O~III, He~I, and OH lines (10$-$20~km/s) \citep{Rey76, wgo01}. This provides further support for the view that the front edge of the Gum Nebula is the actual scattering screen of PSR~B0740$-$28. In Figure \ref{fig:gum}, the arrows representing $\Vec{V}_{\rm scr, \bot}$, $\Vec{V}_{\rm r, \bot}$, and $\vec{V}_{\rm exp, \bot}$ are displayed in white, pink, and yellow, respectively.

In addition, the refractive steering angle ($\theta_{\rm r}$), which reflects the impact of the RISS on each observation, is calculated with the equation $\theta_{\rm r}=(dt/d\nu){\nu}V_{\rm iss}/D$ \citep{Hew80, sw85} and presented in column 12 of Table~\ref{tab:ourresults}. It is evident that $\theta_{\rm r}$ varied more than tenfold within the observed period.

\subsection{The axis of scattering image}
\begin{figure}[H]
\centering
\subfigure{\begin{minipage}{1\linewidth}  
\centering     
\includegraphics[height=5.8cm,width=8cm]{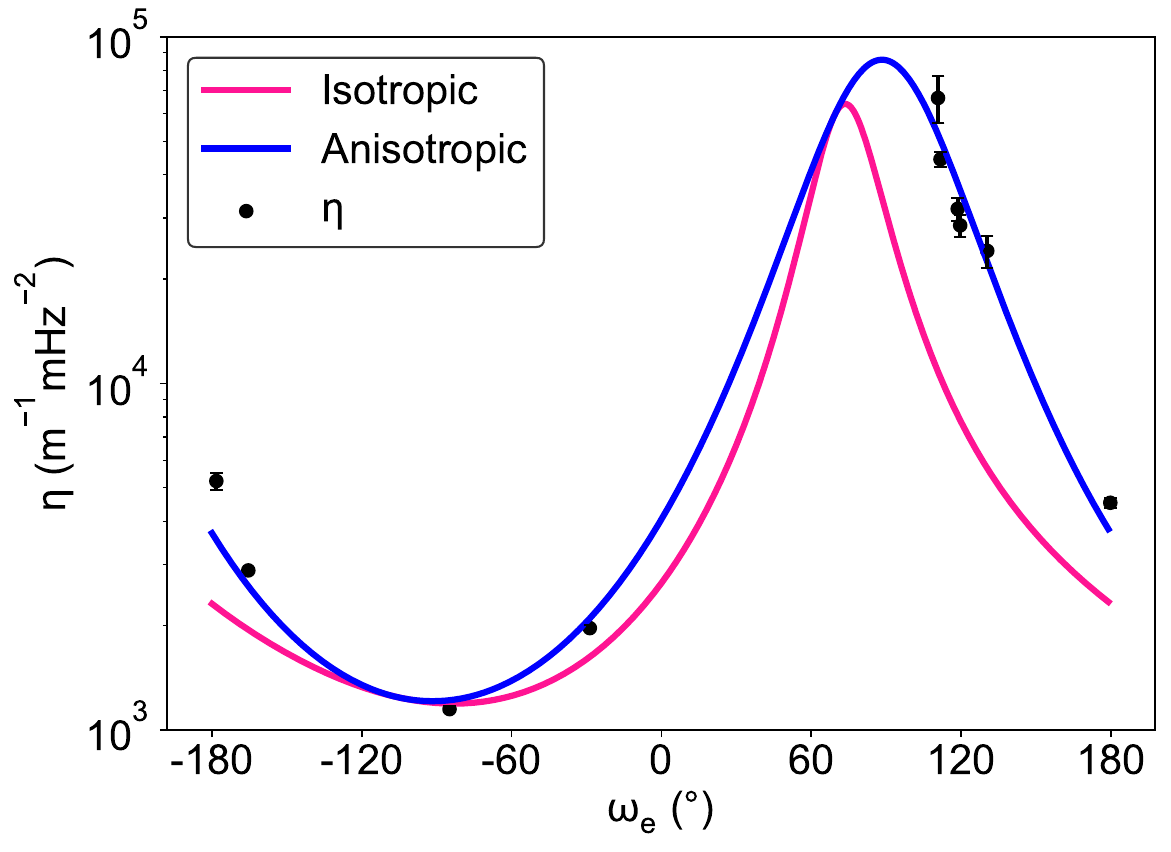}       
\end{minipage}}
\vspace{0.5cm}
\caption{Observed $\eta$ (shown on a logarithmic scale) changing with $\omega_{\rm e}$ and fitted with isotropic scattering (pink curve) and anisotropic scattering (blue curve) models.} 
\label{fig:eta}
\end{figure}
As previously mentioned in Sec.~\ref{sec:secondaryspec}, arc structures have been observed in some of the secondary spectrum plots of 8.60~GHz observations on PSR~B0740$-$28. The degree of arc curvature ($\eta$) depends on the distance to the scattering plasma and the effective transverse velocity of the line of sight relative to the medium ($\Vec{V}_{\rm eff, \bot}$). For an anisotropic thin-screen model, the relationship can be expressed as\citep{crs06}:
\begin{align}
\eta=\frac{Dx/(1+x)^{\rm 2}}{2(\Vec{V}_{\rm eff, \bot}\cdot \vec{a})^2}
\label{equ:etaani}
\end{align}
where $\vec{a}$ is the direction vector of the long axis of the scattering image. This axis physically corresponds to the projection of the turbulence clumps' arranging direction on the plane of the sky \citep{Shi21}. Isotropic scattering can be treated as a special case where $\vec{a}$ is always aligned with $\vec{V}_{\rm eff, \bot}$. $\vec{V}_{\rm eff, \bot}$ is also a vector combination of the transverse velocities of the Earth, the pulsar, and the ISM. This is expressed by the following equation \citep{smc01,crs06}:
\begin{equation}
\Vec{V}_{\rm eff, \bot}=\frac{x}{1+x}\Vec{V}_{\rm p, \bot}+\frac{1}{1+x}\Vec{V}_{\rm earth, \bot}-\Vec{V}_{\rm scr, \bot}
\label{equ:v}
\end{equation}
where $\vec{V}_{\rm eff, \bot}$ can be decomposed into [${V}_{\rm eff, \omega}$, ${V}_{\rm eff, \delta}$] by projecting along the ecliptic longitude and ecliptic latitude, respectively. Let $\psi_{\rm 0}$ be the angle that is clockwise from the ecliptic latitude of the pulsar to the long axis of the scattering image. Then,  Eq.~\ref{equ:etaani} can be expressed as:
\begin{equation}
    \eta=\frac{Dx/(1+x)^{\rm 2}}{2({V}_{\rm eff,\omega}cos\psi_{\rm 0}+{V}_{\rm eff,\delta}sin\psi_{\rm 0})^{\rm 2}}
\label{equ:eta2}
\end{equation}
as $\Vec{a} = (cos\psi_{\rm 0}, sin\psi_{\rm 0})$. Except for $\psi_{\rm 0}$, all parameters related to the location and motion of the screen in Eq.~\ref{equ:eta2} and Eq.~\ref{equ:v} have already been provided in Sec.~\ref{sec:screenfit} from the $V_{\rm iss}$ analysis. In Figure~\ref{fig:eta}, we present the best-fitting curve for $\psi_{\rm 0} = 45.10^{+0.83}_{-0.82}~^{\circ}$ (blue curve), based on $\eta$ results obtained from our 8.60~GHz observations (black points) using the least-squares curve fitting method. It is evident that the anisotropic scattering model provides a superior fit to our results for $\eta$ than the isotropic scattering model prediction (deeppink curve). Moreover, we plot the direction indicator for our best-fitting $\psi_{\rm 0}$ with a lime line in Figure \ref{fig:gum}.

\section{Conclusions}
Utilizing the multi-frequency observing capability of the TMRT, we conducted 21 epochs of simultaneous 2.25~$\&$~8.60~GHz ISS observations on PSR~B0740$-$28 over a period of approximately two years. DISS phenomena were obviously discernible in the dynamic spectrum plots for both the 2.25 and 8.60~GHz observations. In comparison to previous ISS observations, a notable distinction of our study is its simultaneous execution across a broad range of dual frequencies, allowing for more reliable analyses of how the ISS changes with time and frequency. The typical size and regularity of the `scintles' were quantified with $\Delta\nu_{\rm d}$, $\Delta\tau_{\rm d}$, and $dt/d\nu$, which were obtained by fitting the 2D-ACF of the related dynamic spectrum with the 2D-Exponential form function. All ISS parameters showed significant variation with respect to both observational frequency and time. In comparison, the typical values of $\Delta\nu_{\rm d}$ and $\Delta\tau_{\rm d}$ derived from the 2.25~GHz observations were much smaller than those obtained from the 8.60~GHz observations. In addition, we sometimes detected obvious scintillation arc structures in the secondary spectrum plots of our 8.60~GHz observations, while no similar structures were observed in the simultaneous 2.25~GHz observations. The highest frequency record at which arc structures have been observed in the secondary spectrum plots of ISS observations has shifted from 6.656~GHz to 8.60~GHz \citep{xsl23}. Nevertheless, there is no reason why arc structures should not be detectable in the secondary spectra of observations above 8~GHz, other than the declining pulsar flux. The primary benefit of detecting them at 8.60~GHz is the increased frequency lever arm. Benefiting from our simultaneous dual-frequency ISS observations, we obtained more reliable scattering spectral index $\alpha$ values for the ISM from the observation of PSR~B0740$-$28. We found that four epochs of our observations had $\alpha < 4$, providing stronger support for the anomalous scattering reported previously from the ISS perspective. By fitting the annual modulation of $V_{\rm iss, 0}$, we successfully determined the distance to the scattering screen from the observer ($L_{\rm o} = 245^{+69}_{-72}$~pc), which is in good agreement with the distance of the fore-edge of the Gum Nebula. Meanwhile, the best-fitting transverse velocity of the scintillation screen ($|\vec{V}_{\rm scr, \bot}| = 22.4^{+4.1}_{-4.2}$~km/s) is comparable to the vector sum of the Gum Nebula's expanding velocity and its relative velocity to the Sun. Therefore, both $L_{\rm o}$ and $\vec{V}_{\rm scr, \bot}$ are consistent with the relevant parameters of the Gum Nebula, which provides strong evidence for the Gum Nebula acting as the scintillation screen for PSR~B0740$-$28. Furthermore, these results provided additional support for the view that the Gum Nebula acts as the scintillation screen for background pulsars. Compared to the isotropic scattering model, the anisotropic scattering model is a superior explanation for the annual modulation of interstellar scintillation arcs, and the best-fitting orientation of the long axis of the scattering was $\psi_{\rm 0} = 45.10^{+0.83}_{-0.82}~^{\circ}$.

As previously mentioned, 4 of the 21 epochs of our observations showed the scattering spectral index $\alpha < 4.0$, which is classified as anomalous scattering. Anomalous scattering may be induced by breaking one or more of the following preconditions: (i) a single thin, infinite screen or a thick, uniformly distributed ISM from the observer to the target pulsar; (ii) $L_{\rm in}$ and $L_{\rm out}$ of the ISM fluctuations being negligibly small and sufficiently large, respectively; (iii) the ISM fluctuations being isotropic and homogeneous \citep{ldk13}. The anisotropic scattering model is more consistent with the annual modulation of the scintillation arc of our 8.60~GHz observations than the isotropic scattering model. Therefore, we infer that anisotropic scattering likely contributes to the anomalous scattering of PSR~B0740$-$28 \citep{cl01}.

Overall, our ISS observations provide direct evidence for the viewpoint that the Gum Nebula serves as a scattering screen for background pulsars (like PSR~B0740$-$28). On the other hand, changes in both $\alpha$ values and the visibility of scintillation arcs were observed to vary with time, which may be attributed to the complex structure of the Gum Nebula. As most background pulsars of the Gum Nebula are relatively faint, further confirmation of this view should be achieved through long-term ISS observations of these pulsars using larger telescopes. (like the Square Kilometre Array). Furthermore, related observational data will provide more information about the structure of the Gum Nebula and its scattering properties.


\section*{Acknowledgments}
This work was supported in part by the National SKA Program of China (Grant No.~2020SKA0120104), the National Key R~$\&$~D Program of China (Grant No. 2022YFA1603104), and the National Natural Science Foundation of China (Grant Nos. U2031119, 12041301). The image of the Gum Nebula comes from the Southern H-Alpha Sky Survey Atlas (SHASSA), which is supported by the National Science Foundation. We also wish to express our appreciation to Professor Jinlin Han and Chen Wang for their valuable suggestions. Student Han Zhang provided helpful assistance in checking the English expressions in the article. We would also like to express our appreciation to the anonymous referee for their comments and suggestions, which are very useful for improving the manuscript.

\section*{Conflict of interest}
The authors declare that they have no conflict of interest.


\begin{appendix}
\section{Arc fitting in the secondary spectrum}
\label{arc}
To more accurately fit the arc structure, the \texttt{scintools} re-sample the secondary spectrum in the Doppler to transform parabolas into vertical lines, which is done by adjusting the sampling for each row of the spectrum with linear interpolation. After this operation, the secondary spectrum is transformed into the ``normalized'' secondary spectrum. The number of samples decreases with $f_{\lambda}^{\rm 2}$ to get the normalized secondary spectrum $\mathcal{S}(f_{\rm t}/f_{\rm arc},f_{\lambda})$. The arcs will become vertical lines of power in the normalized secondary spectrum. For a reference arc curvature $\eta$, if the arc is located at the normalized $f_{\rm tn} =f_{\rm t}/f_{\rm arc}= 1$, the vertical line at $f_{\rm tn}=f_{\rm b}$ corresponds to an arc with curvature $\eta_{\rm b}=\eta/f_{\rm b}^{\rm 2}$.

By taking cuts of $\mathcal{S}(f_{\rm tn}, f_{\lambda})$ along the $f_{\lambda}$ axis, it reveals that the shape of the power distribution remains approximately constant with $f_{\lambda}$, but the amplitude decays steeply. Thus, with appropriate weighting, $\mathcal{S}(f_{\rm tn}, f_{\lambda})$ can be averaged over $f_{\lambda}$ to improve it and obtain the Doppler profile $D_{\rm t}(f_{\rm tn})$.

To measure $\eta$, \texttt{scintools} rescales the x-axis of the normalized secondary spectrum into physical units of curvature $\eta$ and displays the mean power of the Doppler profile as a function of arc curvature, $P(\eta)$. The peaks observed in $P(\eta)$ correspond to the curvature of each arc.

A more detailed introduction to \texttt{scintools} is provided in the paper \citep{rcb20}. The corresponding plots of the normalized secondary spectrum, the Doppler profile, and the mean power of the Doppler profile as a function of arc curvature are presented in \href{run:.supplement.pdf}{Figure S4} of the online supplementary materials for this paper.

\section{All plots for our observations}
\label{plots}
Our dynamic spectra, 2D-ACFs, secondary spectra, normalized secondary spectra, Doppler profiles, and the mean power of the Doppler profiles can be found in the online supplementary materials for this paper (\href{run:supplement.pdf}{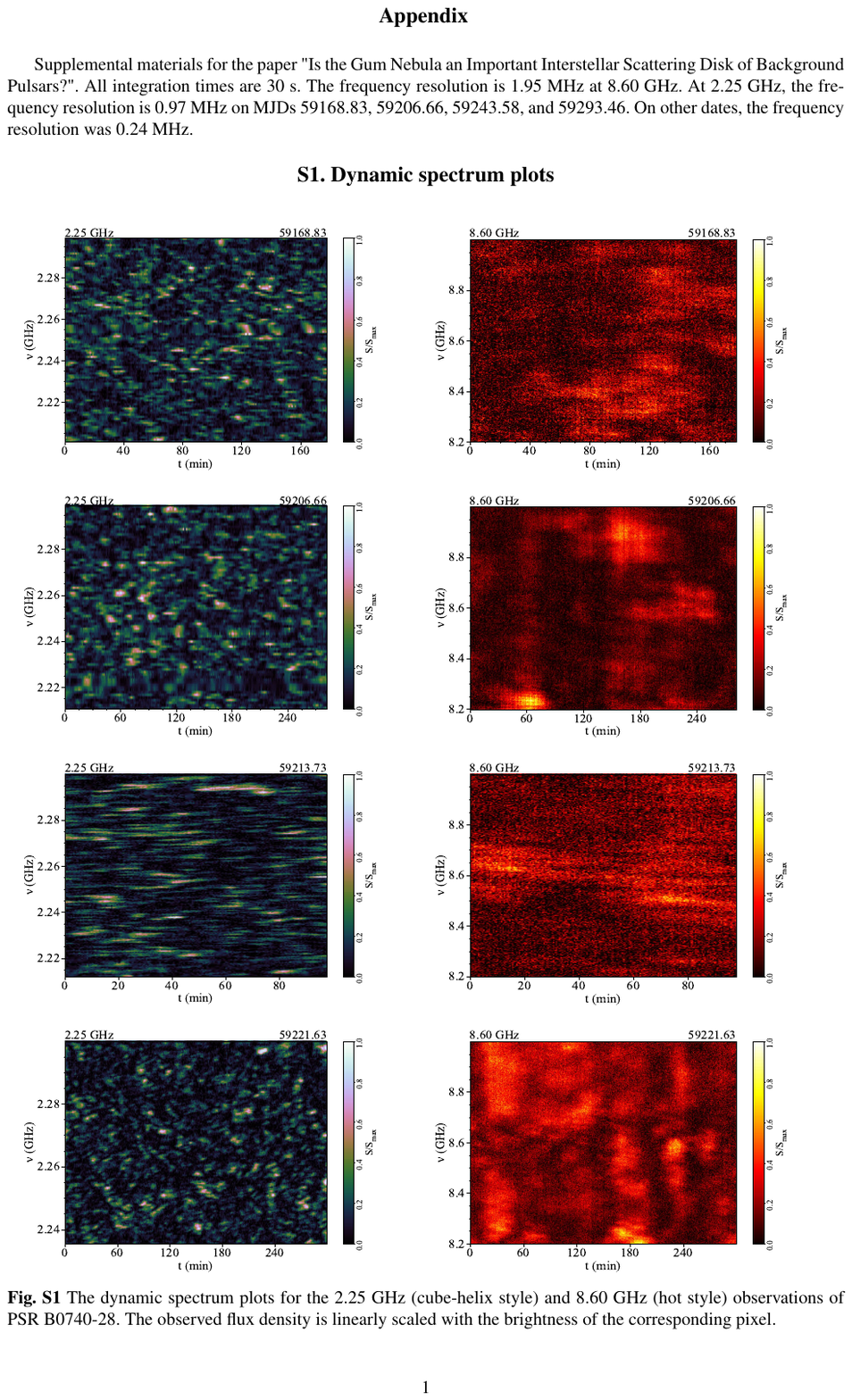}).


\end{appendix}

\newcommand{\ptr}{Phil. Trans. R. Soc. Lond. Ser. A}
\newcommand{\chjas}{Chin. J. Astronom. Astrophys. Suppl.}
\newcommand{\art}{Astronom. Res. Technol.}
\newcommand{\jgrsp}{J. Geophys. Res.: Space Phys.}
\newcommand{\ajp}{Aust. J. Phys.}

\bibliography{b0740}
\bibliographystyle{aasjournal}


\end{document}